\documentclass{elsart}

\journal{JETP Letters}
\pdfoutput=1

\usepackage{graphicx}
\usepackage{bm}
\usepackage{dsfont}
\usepackage{amsmath}
\usepackage{amssymb}
\usepackage{amscd}
\usepackage{float}

\begin{document}

\begin{frontmatter}



\title{Atomic electron shell excitations \\in double-$\beta$ decay}
\author{M.I. Krivoruchenko$^\dagger$, K.S. Tyrin$^\dagger$, F.F. Karpeshin$^*$}
\address{$^\dagger$National Research Centre ``Kurchatov Institute'' \\ Ploschad' Akademika Kurchatova 1$\mathrm{,}$ 123182 Moscow, Russia \\
             $^*$D. I. Mendeleev Institute for Metrology (VNIIM) \\
Moskovsky Ave. 19$\mathrm{,}$ 190005 Saint Petersburg, Russia }

\begin{abstract}
The problem of the transition of electron shells of atoms to excited states in the process of neutrinoless double-$\beta$ decay is investigated.
This subject is crucial for modeling the energy spectrum of $\beta$-electrons, which is sensitive to the mass and Majorana nature of neutrinos.
The dependence of the obtained results on the atomic number indicates an important role of the Feinberg--Migdal effect in the electron shell excitations. We report the overlap amplitudes of the electron shells of the parent atom and the daughter ion for eleven atoms, the two-neutrino double-$\beta$ decay of which was observed experimentally.
In around one-fourth of the cases where the structure of the electron shells is inherited from the parent atom, there is a transition to the ground state or the excited state with the lowest energy. The de-excitation of the daughter ion in the latter scenario is accompanied by the emission of photons in the ultraviolet range, which can serve as an auxiliary signature of double-$\beta$ decay.
The average excitation energy of the electron shells ranges between 300 and 800 eV,
with the variance ranging from $(1.7~\mathrm{keV})^2$ in calcium to $(14~\mathrm{keV})^2$ in uranium.
\end{abstract}

\end{frontmatter}


\newpage

\renewcommand{\theequation}{\arabic{equation}}

Neutrinoless double-$\beta$ decay (0$\nu$2$\beta$) does not preserve the total number of leptons and is particularly 
interesting in the search for departures from the Standard Model (SM).
Similar significance could be found in the quark
sector of SM for processes which violate baryon number conservation, such as proton decay and neutron-antineutron oscillations \cite{Phillips:2016}.
Beyond the SM, any mechanism of 0$\nu$2$\beta$ decay implies the existence
of a Majorana neutrino mass \cite{Schechter:1982,Hirsch:2006}. In
the effective theory, the Majorana neutrino mass, $m_{\nu}$,
is generated by the Weinberg operator of dimension $d = 5$ \cite{Weinberg:1979}.
In the absence of operators of dimension
$d > 5$ and symmetry between left and right elementary fermions, the amplitude of 0$\nu$2$\beta$ decay
with light Majorana neutrino
is proportional to $m_{\nu}$.

Experimental searches for 0$\nu$2$\beta$ decay have been actively preformed for a number of decades.
The GERDA collaboration recently obtained a constraint $m_{\nu} < 0.079 - 0.18$ eV at the confidence
level CL = 90\%
using the isotope $^{76}$Ge \cite{GERDA:2017}.
Similar results
were obtained by the EXO collaboration \cite{Anton:2019} using
xenon-136. Restriction on the neutrino Majorana mass
$m_{\nu}  < 0.3 - 0.9$ eV was also obtained by the collaboration
NEMO-3 using molybdenum-100 \cite{NEMO3:2014}. The SuperNEMO experiment is under
preparation \cite{Arnold:2015}. CUORE's experiments using the isotope $^{130}$Te
\cite{CUORE:2020,CUORE:2022} and KamLAND-Zen with liquid xenon-136 \cite{KamLANDZen:2023}  are in the active phase.

The uncertainty of the upper limit on
the neutrino mass is due to the accuracy of calculations of the nuclear part of the process \cite{Simkovic:2008,Suhonen:2017,Simkovic:2020}.

Experiments to search for 0$\nu$2$\beta$ decay analyse the energy spectrum at the boundary
of the phase space of $\beta$-electrons in order to find a deviation from the
energy spectrum of a more probable two-neutrino double-$\beta$ decay (2$\nu$2$\beta$). Experimentalists
inevitably encounter a problem that has
become widely known in connection with attempts to measure neutrino mass in tritium beta decay:
the daughter atom with a high probability passes into an excited state. This may be the excited
state of a molecule composed of active target atoms. The atoms themselves experience
excitation due to shake-up and shake-off effects
or internal scattering of $\beta$-electrons.
The theory of these processes is developed by Feinberg \cite{Feinberg:1941} and Migdal \cite{Migdal:1941}.
Influence of
these processes on the spectrum of $\beta$-electrons  are especially noticeable near the spectrum boundary.
The effect increases significantly due to the fact that the spread of residual
excitation energies is almost an order of magnitude higher than
the average value \cite{Krivoruchenko:2023}. A similar effect can
be expected from the chemical shift \cite{Lindgren:2004}.

The implications of atom ionization and excitation, first studied in the context of
nuclear physics, are observed in molecular, solid-state systems and are crucial to the experiments
LUX \cite{LUX}, XENON1T \cite{XENON1T}, and
DarkSide-50 \cite{DarkSide:2023}, which are designed to detect dark matter particles.

In double-$\beta$ decays, the daughter ion with a high probability occurs in an excited
state \cite{Krivoruchenko:2020,Karpeshin:2020,Karpeshin:2022,Karpeshin:2023}, which reduces the energy
carried away by $\beta$-electrons.
The energy spectrum of $\beta$-electrons in 0$\nu$2$\beta$ decay is a delta function,
distorted by atomic effects. This peak is considered as the 0$\nu$2$\beta$ decay's signature.
The decay realizes a scenario in which channels with valence electron excitations prevail in probability,
although the average excitation energy, $\mathcal{M}$, and its variance, $\mathcal{D}$,
are essentially saturated by rare electron excitations from inner atomic orbitals.

In this paper, we estimate the deviations of the $\beta$-electrons energy from the decay energy, $Q^*$,
of the 0$\nu$2$\beta$ decay for 11 atoms for which 2$\nu$2$\beta$ decay was experimentally observed.

The binding energy of electrons on the K shell differs from the binding energy of valence electrons by
around three orders of magnitude ($\sim Z^2$) in medium-heavy and heavy atoms, making it difficult to
estimate the magnitude of $\mathcal{M}$ and $\mathcal{D}$ qualitatively.
Excitation of valence electrons with low binding energy obviously dominates the decay probability.
However, the calculations result in unusually large values of average excitation energy and its variance.
Given that the accuracy of calculations in multi-particle problems is also limited, the paper considers
several approaches, including the Thomas-Fermi (TF) model \cite{Landau:1987}, the Thomas-Fermi-Dirac-Weizsacker model (TFDW) \cite{Dirac:1930,Weizsacker:1935,Kirzhnits:1967,Gross:1979,Stich:1982},
non-relativistic Rutaan-Hartree-Fock (RHF) formalism \cite{Clementi:1974} and relativistic Dirac-Hartree-
Fock formalism (DHF) \cite{Lu:1971,Desclaux:1973,HUANG:1976,Dyall:1989,Grant:2008}.
When the outcomes are compared, the magnitude of uncertainty in the parameters of interest can be evaluated.

Each of these approaches has its advantages and limitations. Unlike the TF model, in the TFDW model
the electron density is finite at the nucleus, which makes it possible to determine the variance within
the model. In the RHF method, the wave functions of orbitals are parametrized analytically, which makes
it possible to find exchange contributions to the variance and other observables, but the applicability
of the method is restricted to light and medium-heavy atoms. Within the framework of DHF,
the basic
properties of atomic electron shells are tabulated in \cite{Lu:1971,Desclaux:1973,HUANG:1976}
and implemented in the form of software packages such as G\textsc{rasp}-2018
\cite{Dyall:1989,Grant:2008} and RAINE \cite{Band:1977,Band:2002}.

In what follows, the system of atomic units $\hbar = m = e = 1$, $c = 137$ is used, where $m$ is the electron mass, $e$ is the proton
charge, $c$ is the speed of light. Let $\hat{H}_{Z,N}$ be the Hamiltonian of $N$ electrons of an ion with
a nucleus charge $Z$. We denote $|Z,N\rangle$ the ground
state and
$E_{Z,N}$ the binding energy of the electrons, so that $\hat{H}_{Z,N}|Z,N\rangle = E_{Z,N}|Z,N\rangle$.


\begin{table} [!t]
\addtolength{\tabcolsep}{-5pt}
\renewcommand{\arraystretch}{1.2}
\scriptsize
\centering
\caption{
The average excitation energy of the electron shells of the daughter ion and the variance
for eleven atoms, the $2\nu2\beta$ decay of which was observed experimentally.
The second column contains the values
of the mass difference, $Q$, of the neutral atoms involved in the decay.
The fourth column shows the overlap integrals of the electron shells of the parent atom and the twice ionized daughter atom,
calculated with the use of the software package G\textsc{rasp}-2018 \cite{Dyall:1989,Grant:2008}.
The average energy of the electron shell excitations
of the daughter ion is shown in the TF, TFDW and DHF models, the upper bound of the variance $\bar{\mathcal{D}}$ is shown in the TF, TFDW models,
and the variance $\mathcal{D}$ -- in the DHF and RHF models. The values of $\mathcal{M}_{\mathrm{DHF}}$ and $\mathcal{D}_{\mathrm{DHF/a}}$
excluding exchange contributions
are obtained using the results of \cite{HUANG:1976} and \cite{Desclaux:1973}, respectively.
$\mathcal{D}_{\mathrm{DHF/b}}$ includes exchange contributions.
To calculate $\mathcal{D}_{\mathrm{RHF/a}}$ without and $\mathcal{D}_{\mathrm{RHF/b}}$ with exchange contributions, the wave
functions of orbitals in the RHF method are used \cite{Clementi:1974}.
The double ionization energy $I_2$ \cite{Kramida:2022} is rounded to three significant digits.
The predictions of the non-relativistic models TF, TFDW and RHF are limited by the values of the nuclear charge $Z\le 54$.
\label{main:tab}
}
\vspace{2mm}
\begin{tabular}{|rrr@{\hspace{2mm}}lc@{\hspace{2mm}}cc@{\hspace{-1mm}}c@{\hspace{-1mm}}c@{\hspace{-1mm}}c@{\hspace{-0mm}}c@{\hspace{-1mm}}c@{\hspace{-1mm}}
c@{\hspace{-1mm}}c@{\hspace{-1mm}}c@{\hspace{-1mm}}c@{\hspace{-1mm}}|}
\hline
\hline
\multicolumn{3}{|c}{Process}
&
$\begin{array}{c}
Q \\
\textup{[keV]}
\end{array}$
&
Ref.
&
$K_Z$
&
$\begin{array}{c}
\mathcal{M}_{\mathrm{TF}} \\
\textup{[eV]}
\end{array}$
&
$\begin{array}{c}
\mathcal{M}_{\mathrm{TFDW}}  \\
\textup{[eV]}
\end{array}$
&
$\begin{array}{c}
\mathcal{M}_{\mathrm{DHF}}  \\
\textup{[eV]}
\end{array}$
&
$\begin{array}{c}
I_2  \\
\textup{[eV]}
\end{array}$
&
$\begin{array}{c}
\bar{\mathcal{D}}_{\mathrm{TF}}^{1/2} \\
\textup{[keV]}
\end{array}$
&
$\begin{array}{c}
\bar{\mathcal{D}}_{\mathrm{TFDW}}^{1/2} \\
\textup{[keV]}
\end{array}$
&
$\begin{array}{c}
\mathcal{D}_{\mathrm{DHF/a}}^{1/2} \\
\textup{[keV]}
\end{array}$
&
$\begin{array}{c}
\mathcal{D}_{\mathrm{DHF/b}}^{1/2} \\
\textup{[keV]}
\end{array}$
&
$\begin{array}{c}
\mathcal{D}_{\mathrm{RHF/a}}^{1/2} \\
\textup{[keV]}
\end{array}$
&
$\begin{array}{c}
\mathcal{D}_{\mathrm{RHF/b}}^{1/2} \\
\textup{[keV]}
\end{array}$
\\
\hline
$_{\;\,20}^{\;\,48}\mathrm{Ca} $ & $ \rightarrow$ & $ _{\;\,22}^{\;\,48}\mathrm{Ti}$   & 4267.98(32)  & \cite{Kwiatkowski:2014} &  0.466   & 335  & 247 & 299 & 20.4& 1.25 &  2.43  &1.70 &1.65 & 1.66  & 1.61  \\
$_{\;\,32}^{\;\,76}\mathrm{Ge} $ & $ \rightarrow$ & $ _{\;\,34}^{\;\,76}\mathrm{Se}$   & 2039.006(50) & \cite{Suhonen:2007}     &0.575& 383  & 246 & 369 & 30.9& 2.16 &  3.92  & 2.88 &2.77& 2.72  & 2.62  \\
   &&                                                         & 2039.061(7)  & \cite{GERDA:2017}         &      &      &     &     &      &        &&      &&       &       \\
$_{\;\,34}^{\;\,82} \mathrm{Se} \,$ & $ \rightarrow$ & $ _{\;\,36}^{\;\,82}\mathrm{Kr}$  & 2997.9(3)    & \cite{Lincoln:2013}     &0.597& 384  & 238 & 377 &38.4& 2.31 &  4.17  &3.09  &2.97& 2.90  & 2.79\\
$_{\;\,40}^{\;\,96} \mathrm{Zr} \,$ & $ \rightarrow$ & $ _{\;\,42}^{\;\,96}\mathrm{Mo}$ & 3356.097(86) & \cite{Alanssari:2016}    &0.518& 422  & 246 & 409 &23.3& 2.78 &  4.92  & 3.76 &3.60& 3.44  & 3.29\\
$_{\;\,42}^{100}\mathrm{Mo} $ & $ \rightarrow$ & $ _{\;\,44}^{100}\mathrm{Ru}$ & 3034.40(17)  & \cite{Rahaman:2008}     &0.564& 428  & 241 & 419 &24.1& 2.94 &  5.17  & 4.00 &3.82& 3.62  & 3.46\\
$_{\;\,48}^{116}\mathrm{Cd} $ & $ \rightarrow$ & $ _{\;\,50}^{116}\mathrm{Sn}$ & 2813.50(13)  & \cite{Rahaman:2011}     &0.601& 451  & 229 & 442 &22.0& 3.42 &  5.92  & 4.74 &4.51& 4.17  & 3.97\\
$_{\;\,52}^{128}\mathrm{Te} \,$ & $ \rightarrow$ & $ _{\;\,54}^{128}\mathrm{Xe}$ & 865.87(131)  & \cite{Scielzo:2009}     &0.589& 452  & 206 & 457 &33.1& 3.74 &  6.42  & 5.29 &5.04& 4.53  & 4.32\\
$_{\;\,52}^{130}\mathrm{Te} \,$ & $ \rightarrow$ & $ _{\;\,54}^{130}\mathrm{Xe}$ & 2526.97(23)  & \cite{Rahaman:2011}     &0.589& 452  & 206 & 457 &33.1& 3.74 &  6.42  & 5.29 &5.04& 4.53  & 4.32\\
$_{\;\,54}^{136}\mathrm{Xe} $ & $ \rightarrow$ & $  _{\;\,56}^{136}\mathrm{Ba}$ & 2457.83(37)  & \cite{Redshaw:2007}     &0.606& 476  & 217 & 465 &15.2& 3.91 &  6.67  & 5.57 &5.31& 4.71  & 4.49\\
$_{\;\,60}^{150}\mathrm{Nd} $ & $ \rightarrow$ & $ _{\;\,62}^{150}\mathrm{Sm}$ & 3371.38(20)  & \cite{Kolhinen:2010}    &0.519&      &     & 514 &16.7&      &        & 6.50 &6.20&   &         \\
$_{\;\,92}^{238}\mathrm{U} \;\;$ & $ \rightarrow$ & $ _{\;\,94}^{238}\mathrm{Pu}$ & 1437.3       & \cite{Firestone:1996}   &0.546&      &     & 774 &17.5&      &        & 14.58 &13.90&   &        \\
\hline
\hline
\end{tabular}%

\end{table}


The Hamiltonian of the daughter ion's electrons is related to the Hamiltonian of the parent neutral atom's electrons via the relation
\begin{equation} \label{decompose}
\hat{H}_{Z+2,Z} = \hat{H}_{Z,Z} -2 \sum_{i}\frac{1}{{r}_{i}},
\end{equation}
\noindent
where $r_i = |\mathbf{r}_{i}|$, $\mathbf{r}_{i}$ is the coordinate of the $i$th electron,
and summation is performed by $i=1,\ldots,Z$.
The electrons of the daughter ion are in the state $|Z,Z\rangle$ for the next moment after the decay,
while the nucleus acquires a charge of $Z+2$. The relationship
\begin{equation}
\mathcal{M} = \langle Z,Z|\hat{H}_{Z+2,Z}|Z,Z\rangle - \langle Z + 2,Z|\hat{H}_{Z+2,Z}|Z + 2,Z\rangle \label{defM}
\end{equation}
determines the average excitation energy of the daughter ion's electrons,
or, with account of Eq.~(\ref{decompose}),
\begin{equation} \label{compose}
\mathcal{M} =  E_{Z,Z} + 2 Z^{-1}E_{Z,Z}^{\mathrm{C}} - E_{Z+2,Z},
\end{equation}
where $E_{Z,N}^{\mathrm{C}}$ is the Coulomb interaction energy of the electrons with the nucleus.

Table 1 shows the results of the calculation of the excitation energy in the TF, TFDW and DHF models.
First, the values $\mathcal{M}^{\prime}$ are found, which
differ from $\mathcal{M}$ by replacing in Eq.~(\ref{compose}) the binding energy of the electrons
of the ion $E_{Z+2,Z}$ with the binding energy of the electrons of the neutral atom $E_{Z+2,Z+2}$.
The difference between $E_{Z+2,Z}$ and $E_{Z+2,Z+2}$ is equal to the double ionization energy, $I_2$;
there is a relation $\mathcal{M} = \mathcal{M}^{\prime} - I_2$.
The experimental values of $I_2$ are collected in \cite{Kramida:2022}.

In the TF model, the calculations are carried out according to the scheme of \cite{Krivoruchenko:2023}.
The TFWD model, being a generalization of the TF model, additionally takes into account exchange contribution
to the energy of electron gas \cite{Dirac:1930} and spatial inhomogeneity in the electron density \cite{Weizsacker:1935}.
A consistent semiclassical decomposition of the density functional with account of the inhomogeneity
can be found
in the monograph by Kirzhnits \cite{Kirzhnits:1967}.
In its simplest form
\begin{eqnarray}
E_{Z,N} &=& \int d\mathbf{r} \left( - \frac{Z}{r}n(\mathbf{r}) + c_1 n^{5/3}(\mathbf{r}) + c_2 n^{4/3}(\mathbf{r}) + c_3 \frac{(\nabla n(\mathbf{r}))^2}{n(\mathbf{r})}\right)  \\
    &~&~~~~~~~~~~~~~~~+
    \frac{1}{2}\int d\mathbf{r}d\mathbf{r}^{\prime}n(\mathbf{r})\frac{1}{|\mathbf{r} - \mathbf{r}^{\prime}|}n(\mathbf{r}^{\prime}). \nonumber
\end{eqnarray}
Here, $n(\mathbf{r})$ is the electron density, the first term under the integral sign represents the interaction energy of the electrons
with the nuclus, $E_{Z,N}^{\textrm{C}}$, the second term is the kinetic energy, the third one is the exchange energy,
the fourth one is the Weizsacker gradient correction \cite{Weizsacker:1935}. The last term is the interaction energy
of electrons. The coefficients $c_i$ equal
\begin{equation}
c_1 = \frac{3}{10}(3\pi^2)^{2/3},~~c_2 = - \frac{3}{4}\left(\frac{3}{\pi}\right)^{1/3},~~c_3 = \frac{\lambda}{8}.
\end{equation}
The value $\lambda = 1/5$ of the phenomenological models \cite{Gross:1979,Stich:1982} is used.

In \cite{Gross:1979}, the binding energy of neutral atoms N, Ne, Ar, Kr, Xe
with filled valence shells is calculated using the TFDW model. Parameterization of the results gives
$E_{Z,Z} = - 0.536 Z^{2.38}$, which is not much different from
the TF model, where $E_{Z,Z} = - 0.764 Z^{7/3}$. The energy of the Coulomb interaction of electrons
with the nucleus is calculated using the screening function. Integrating the expression for
$E_{Z,Z}^{\textrm{C}}$ by parts,
the action of the Laplacian is transferred to the Coulomb potential, which gives a delta function at the origin.
The difference between the total potential and the nuclear potential occurs as a multiplier.
The interaction energy turns out to be $Z^2(a - b)$, the screening
function parameters $a$ and $b$ are given in Table II of \cite{Gross:1979}.
The fitting gives $E_{Z,Z}^{\textrm{C}} = - 1.270 Z^{2.38}$ in agreement with the virial theorem.
The parameterization accuracy is not worse than 0.5\%.
The corresponding results for $\mathcal{M}$ are shown in Table 1.

The average values of $\langle Z,Z|r^{-1}_i|Z,Z\rangle$ required to estimate $E_{Z,Z}^{\mathrm{C}}$ in the DHF method
are tabulated in \cite{Clementi:1974,Lu:1971,Desclaux:1973,HUANG:1976}. In \cite{HUANG:1976}, the values of
$E_{Z,Z}^{\mathrm{C}}$ are also provided. The results of calculations of $\mathcal{M}$ within the framework of
DHF model \cite{HUANG:1976} are shown in Table 1.

The TF and DHF models agree well with each other and agree qualitatively with the predictions of the TFDW model.

The variance of the electron excitation energy is determined by the formula
\begin{equation}
\mathcal{D} =  \langle Z,Z|\hat{H}_{Z+2}^2|Z,Z\rangle - \langle Z,Z|\hat{H}_{Z+2}|Z,Z\rangle^2.  \label{disp}
\end{equation}
Taking into account Eq.~(\ref{decompose}) we have
\begin{equation}
\frac{1}{4} \mathcal{D} =  \sum_{ij}\langle Z,Z|\frac{1}{r_i} \frac{1}{r_j} |Z,Z\rangle -
\langle Z,Z|  \sum_{i} \frac{1}{r_i} |Z,Z\rangle ^2.  \label{disp}
\end{equation}
The summation is performed in the range $1 \leq i,j\leq Z$. In the TF and TFDW models,
the two-particle electron density is not defined, however, it is possible to fix
the upper limit of the variance \cite{Krivoruchenko:2023}:
\begin{equation} \label{ineq}
\frac{1}{4} \bar{\mathcal{D}} = \int d\mathbf{r} \frac{1}{{r}^2} n(\mathbf{r}) - Z^{-1} \left( \int d\mathbf{r} \frac{1}{{r}} n(\mathbf{r}) \right)^2.
\end{equation}
Calculation of the integral of $1/r^2$ over the electron density distribution
in the TFDW model leads to values that can be parameterized as
\begin{equation} \label{disp1}
\int d\mathbf{r} \frac{1}{r^2} n(\mathbf{r}) = 5.81 Z^{2.00}.
\end{equation}
The parameterization accuracy is not worse than 5\%.
The values of $\bar{\mathcal{D}}$ in the TF and TFDW models are
shown in Table 1.

In the DHF method, it is possible to estimate not only the upper bound of the variance, but also the variance itself.
In disregard of exchange effects
$\mathcal{D}$ is calculated from Eq.~(\ref{disp}) after factorization of the average value under
the double summation sign. The corresponding results, using the tabulated values of averages $1/r_i$ and $1/r^2_i$
for the electron orbitals \cite{Desclaux:1973}, are shown in Table 1.

The exchange effects are taken into account by averaging the two-particle operator
over the total wave function of electrons of the atom. In the
one-determinant approximation, the wave function has the form
\begin{equation}
\Psi _{\alpha_1 \alpha_2 \ldots \alpha_N}=\frac{1}{\sqrt{N!}}\epsilon
^{s_{1}s_{2}...s_{N}}\phi _{\alpha _{s_{1}}}^{1}\phi _{\alpha
_{s_{2}}}^{2} \ldots \phi _{\alpha _{s_{N}}}^{N},  \label{slater}
\end{equation}
where $\phi_{\alpha}^{i}$ are the wave functions of electrons, the index $i=1,...,N$ counts the spatial
coordinates and spin indices, the index $\alpha$ counts the quantum numbers of orbitals.
In the case under consideration, $\alpha=(njlm)$, where $n$ is the principal quantum number,
$j$ is the total angular momentum, $m$ is its projection,
$l=j\pm 1/2$ is the orbital angular momentum.
A fixed set of quantum numbers $(\alpha_1, \alpha_2, \ldots, \alpha_N)$ determines the state of the electron
shells of the atom. The tensor $\epsilon^{s_1 s_2\ldots s_{N}}= \pm 1$ performs antisymmetrization.

The functions $\phi_{\alpha}^{i}$ are orthonormal. We write them as the product of the radial and angular parts:
\begin{equation} \label{deco}
\phi _{njlm}^{i}=R_{njl}(r_i)\Omega
_{jm}^{l}(\mathbf{n}_i).
\end{equation}
Here $R_{njl}(r)$ is a real function, $\Omega_{jm}^{l}(\mathbf{n})$ is a spherical spinor depending on the unit vector $\mathbf{n} = \mathbf{r}/|\mathbf{r}|$.
We denote by $\kappa_{njl}$ the number of occupied energy levels with quantum numbers $(njl)$.
In the case of fully occupied energy levels, as well as cases allowing for each pair of $(jl)$ the existence of no more than one partially
occupied energy level with the maximum total angular momentum, $j^{\max} = \kappa_{njl}(2 j + 1 - \kappa_{njl})/2$,
one can simplify Eq.~(\ref{disp}) by replacing the summation over
electrons by the summation over energy levels:
\begin{equation}
\frac{1}{4}\mathcal{D} = \sum_{njl} \kappa_{njl} \langle njl|r^{-2}|njl\rangle
- \sum_{nn^{\prime }jl} \min(\kappa_{njl},\kappa_{n^{\prime}jl}) \langle njl|r^{-1}|n^{\prime }jl\rangle^{2}. \label{dispexchange}
\end{equation}
The matrix elements are defined according to
\begin{equation}
\langle njl|h(r)|n^{\prime }jl \rangle =\int
r^{2}drh(r) R_{_{njl}}(r)R_{_{n^{\prime }jl}}(r).
\end{equation}

The sum of the diagonal components $n = n^{\prime}$ of Eq. (\ref{dispexchange}) coincides with the right
side of Eq.~(\ref{disp}) through factorization of the mean value under the sign of the double sum,
as it is assumed in the TF and TFDW estimates. The components with $n\neq n^{\prime}$ in the second term of
Eq. (\ref{dispexchange}) are related to exchange effects. The exchange effects reduce the variance.

In the RHF method, the functions $R_{njl}(r)$ are tabulated \cite{Clementi:1974}.
To calculate the variance taking into account exchange effects, knowledge of the off-diagonal matrix
elements $\langle njl|r^{-1}|n^{\prime}jl\rangle$ and $\langle njl|r^{-2}|n^{\prime}jl\rangle$ is required. Table 2 shows
the results for molybdenum atom in the RHF method. The diagonal matrix elements are compared
with those in the DHF method \cite{Desclaux:1973}. There is some systematic underestimation of the average values
in comparison with the DHF method, which is due to the shift in relativistic models of the electron density
to smaller distances \cite{Grant:2008}.
A similar pattern holds for other 10 atoms.
Accordingly,
the variance in the RHF method without taking into account exchange contributions is also systematically lower than the predictions of the DHF method.

The average values of $1/r$ and $1/r^2$ in inner and outer orbitals are approximately in the ratios $Z:1$ and $Z^2:1$,
which is consistent with the values of the diagonal matrix elements in Table 2. In cases where for a partially occupied level the total
angular momentum is not the maximum and/or where there exist more than one partially occupied level for a pair of $(jl)$,
the formula (\ref{dispexchange}) is used as an approximation. Since in medium-heavy and heavy atoms,
the main contribution to the variance comes from electrons in inner shells, where $\kappa_{njl} = 2j+1$,
one can expect that accuracy of such an estimate is quite high.

The results of calculations of $\mathcal{D}$ in the RHF method, taking into account exchange contributions,
are presented in Table 1. For comparison, the results of calculations without exchange effect are also provided.
The agreement with the TF, TFDW, RHF and DHF models is quite satisfactory.

For applications, we recommend the values of excitation energy $\mathcal{M}_{\mathrm{DHF}}$,
for variance -- $\mathcal{D}_{\mathrm{DHF/b}}$, as theoretically the most justified.
The estimate of $\mathcal{D}_{\mathrm{DHF/b}}$ differs from $\mathcal{D}_{\mathrm{DHF/a}}$ in that
it includes exchange corrections found by the RHF method.
Taking into account various approximations, the uncertainty in
$\mathcal{M}_{\mathrm{DHF}}$ and $\mathcal{D}_{\mathrm{DHF/b}}$ can be estimated at $< 10\%$.

In the non-relativistic TF, TFDW and RHF models
$\mathcal{M}$ only weakly depends on $Z$, while $\mathcal{D}$ grows approximately as $Z^2$.
This behavior is perfectly consistent with the highlighted role of K electrons, whose nonrelativistic excitation theory
in $\beta$ decay is developed in \cite{Feinberg:1941,Migdal:1941}.




\begin{table}[t]
\addtolength{\tabcolsep}{-1.2 pt}
\renewcommand{\arraystretch}{1.2}
\scriptsize
\centering
\caption{Matrix elements $\langle nl|r^{-1}|n^{\prime} jl\rangle$ and $\langle n jl|r^{-2}|n^{\prime} jl\rangle$
for $n\leq n^{\prime}$ electron orbitals in a molybdenum atom. Calculations use electron wave functions of the RHF method~\cite{Clementi:1974}
with degeneracy in $j$.
In the lower part of table, diagonal matrix elements $n=n^{\prime}$ of the relativistic DHF method~\cite{Desclaux:1973} are given;
the upper and lower rows of P and D waves correspond to $j=l +  1/2$ and $j=l - 1/2$, respectively.
}
\vspace{2mm}
\label{tab:table40}
\begin{tabular}{|c|}
\hline
\hline
$_{42}$Mo \\
\begin{tabular}{c|rrrrr|r|rrr|r|rr}
\hline
$\langle njl |r^{-1}|n^{\prime}jl\rangle$ & 1S      & 2S      & 3S     & 4S      & 5S~     &    & 2P      & 3P     & 4P~     &    & 3D          & 4D~ \\
\hline
1S & 41.49   & 7.962   &  3.231 & -1.255 & 0.321~  &      &         &        &           &      &             &              \\
2S &         & 9.378   &  2.160 & -0.803 & 0.204~  & ~2P~ &  9.339  & -1.858 & -0.626~   &      &             &              \\
3S &         &         &  3.264 & -0.665 & 0.163~  & ~3P~ &         &  3.164 &  0.582~   & ~3D~ & 2.970       & -0.361~       \\
4S &         &         &        &  1.171 &-0.149~  & ~4P~ &         &        &  1.052~   & ~4D~ &             &  0.714~       \\
5S &         &         &        &        & 0.327   &      &         &        &           &      &             &              \\
\hline
\cite{Desclaux:1973}& 43.55 &  9.939 & 3.409  & 1.209  &   0.322~  &    & 9.412 & 3.190 &  1.059~ &    & 2.958     & 0.695~       \\
&         &         &        &         &         &                      & 9.879 & 3.300 &  1.089~ &    & 2.987     & 0.705~       \\
\hline
$\langle njl |r^{-2}|n^{\prime}jl\rangle$ & 1S      & 2S      & 3S     & 4S      & 5S~      &    & 2P      & 3P     & 4P~      &    & 3D          & 4D~           \\
\hline
1S & 3455.   &  984.9  & 410.3  & -160.1   & 40.94~   &      &         &        &          &    &             &              \\
2S &         &  357.8  & 141.7  & -55.02  & 14.06~   & ~2P~ & 118.4   & -37.69 & -13.17~   &    &             &              \\
3S &         &         &  65.20 & -24.42  &  6.223~  & ~3P~ &         &  21.34 &   6.697~  & ~3D~ & 11.17      & -2.120~       \\
4S &         &         &         &   10.41 & -2.564~  & ~4P~ &         &        &   3.157~  & ~4D~ &             & 0.965~       \\
5S &         &         &         &         &  0.748~  &      &         &        &          &    &             &              \\
\hline
\cite{Desclaux:1973}& 4005. & 439.4 & 80.03  &   12.73 &  0.830~ &    & 120.7  & 21.93 & 3.243~ &    & 11.11      & 0.930~       \\
&         &         &        &         &         &                     & 141.5  & 25.50 & 3.744~ &    & 11.37      & 0.960~
\end{tabular}\\
\hline
\hline
\end{tabular}
\end{table}

The parameter $K_Z$ given in Table 1 represents the overlap amplitude of the wave functions of all the electrons
in the ground state of the parent atom with the wave functions of the electrons in the ground state  of the twice ionized daughter atom, whose electrons have retained their initial configuration.
The corresponding wave functions of the electrons are not orthogonal because the charges of the nuclei before and after shaking
differ by two units, and as a result, the overlap of electron wave functions with the identical quantum numbers is not equal
to one. The daughter ion gets excited as a result.
The value $K^2_Z$ determines the probability of inheriting quantum numbers by the electrons
and, accordingly, the absence of shaking effects.

To estimate $K_Z$, a multiparticle calculation using the DHF method is required, which was performed
using the software package G\textsc{rasp}-2018 \cite{Dyall:1989,Grant:2008}.
A set of large $f_{njl}^{+}(r)$ and small $f_{njl}^{-}(r)$
radial components of electron wave functions
for all quantum numbers $(njl)$
is obtained for each parent atom of Table 1 with an appropriate electron configuration and a total
angular momentum corresponding to the ground state of the parent atom's electrons.
Similarly, for a daughter ion with a nuclear charge $Z+2$, a
set of radial components $\Tilde{f}_{njl}^{\pm}(r)$ is obtained. The overlap amplitude $\mathcal{O}_{njl}$
of the wave functions of electrons with the same quantum numbers is equal to
\begin{equation*}
\mathcal{O}_{njl}= \int \left(\Tilde{f}_{njl}^{+}(r)f_{njl}^{+}(r)+\Tilde{f}_{njl}^{-}(r)f_{njl}^{-}(r) \right) r^{2}dr.
\end{equation*}
In the one-determinant approximation and without taking into account exchange terms, the amplitude of $K_Z$ is equal
to the product of the amplitudes of $\mathcal{O}_{njl}$ in the degree equal to the occupation number of the corresponding level:
\begin{equation}
K_Z=\prod_{njl} \left( \mathcal{O}_{njl}\right)^{\kappa_{njl}}.
\end{equation}

For a wide range of atomic numbers $Z$, the values of $K_Z$ turn out to be close to 1/2. The probability of $K^2_Z \sim 1/4$ is quite small,
which indicates the dominance of channels with the excited electron shells of atoms
in agreement with the phenomenological analysis \cite{Krivoruchenko:2020}.

The above approach assumes that the resulting configuration of the daughter ion is the ground state with quantum
numbers of electrons inherited from the parent atom.
However spectroscopic analysis shows that this condition is not always met.
For example, the Ti III ion formed after $0\nu2\beta$ decay of Ca with the electron configuration [Ar]4s$^2$,
in the ground state has the configuration [Ar]3d$^2$. Strictly speaking,
this fact means that the overlap is exactly zero: $K_Z \equiv 0$, therefore, the decay with the dominant probability
is accompanied by the excitation of the electron
shells of the atom. The energy of the [Ar]4s$^2$ lowest configuration exceeds that of the [Ar]3d$^2$ configuration of Ti III by 12.7 eV.
A similar situation occurs in double-$\beta$ decay
atoms of Zr, Mo, Nd and U. In these cases, the amplitude $K_Z$, given
in Table 1, is the amplitude of the transition to the most likely excited state of the electron shells
of the daughter ion. In approximately every fourth case, double-$\beta$ decay of Ca, Zr, Mo, Nd and U
is accompanied by the de-excitation of the electron shells of atoms from the unique excited state to the ground state with
the emission of a series of photons of the ultraviolet range.
The observation of these photons, whose wavelengths are well known, can serve as an auxiliary signature for the identification of decay.


Knowledge of the parameters $K_Z$, $\mathcal{M}$ and $\mathcal{D}$ is sufficient to construct simple models
of the energy distribution of $\beta$-electrons in $0\nu2\beta$ decay.
With a probability of $K_Z^2$, the electrons
of the decaying atom remain in the lowest energy state, preserving their quantum numbers,
with a probability of $1 - K_Z^2$ they pass into an excited state.
The conditional probability of transition to an excited state with energy $\epsilon$ in the interval $d\epsilon$
is denoted by $w(\epsilon/Q^*)d\epsilon/Q^*$. The total probability density takes the form
\begin{equation}
p(\epsilon) = K_Z^2\delta(\epsilon) + (1 - K_Z^2)w(\epsilon/Q^{*})/Q^{*}.
\end{equation}
The binomial distribution is used for $w(x)$,
which has a certain universatility and is widely used in modeling random processes \cite{Korolyuk:1985}.
The distribution has two free parameters, which are fixed by normalization to the average
value of $\mathcal{M} = \int_0^{Q^{*}} d\epsilon \epsilon p(\epsilon)$ and the mean square of the energy
$\mathcal{D}+ \mathcal{M}^2 =\int_0^{Q^{*}} d\epsilon \epsilon^2 p(\epsilon)$.

Based on the DHF model predictions, we calculate the maximum deviation, $\Delta T_{\max}$,
of the $\beta$-electrons  energy from the decay energy $Q^{*}$. $\Delta T_{\max}$ can be
determined through the equation
\begin{equation*}
p_T = \int_0^{\Delta T_{\max}} d\epsilon p(\epsilon)
\end{equation*}
for a given probability, $p_T$.
The value of $p_T = 0.9$ corresponds to the
deviations of the $\beta$-electrons energy from $Q^*$ less than $\Delta T_{\max} =  180$ eV (Ca),
18 eV (Ge), 19 eV (Se), and $\Delta T_{\max} < 5$ eV for Zr, Mo, Cd, Te, Xe, Nd, U.
At the probability of $p_T = 0.95$,
the deviations do not exceed $\Delta T_{\max} = 1.16$ keV (Ca), 0.55 keV (Ge), 0.44 keV (Se), 0.25 keV (Zr), 0.18 keV (Mo),
69 eV (Cd), 22 eV ($^{128}$Te), 30 eV ($^{130}$Te), 19 eV (Xe), 11 eV (Nd), and $\Delta T_{\max} < 5$ eV for U.

The decay energy without the energy taken away by neutrinos is measured in calorimetric detectors,
where the energy resolution reaches a few keV (GERDA).
Using a track calorimeter, the SuperNEMO experiment measures the energy of $\beta$-electrons in $0\nu2\beta$-selenium decay
with an uncertainty of 4\%
at an energy of $Q$.
Innovative technologies with excellent energy resolution are in high demand for reducing background noise
and for tracking the impact of atomic shell excitations on the neutrino mass constraints.



To summarize, the overlap amplitudes for the electron shells of the parent atom and the daughter ion for each atom whose $2\nu2\beta$ decay was observed experimentally were found.
In the double-$\beta$ decay of atoms $^{82}$Se, $^{96}$Zr, $^{100}$Mo, $^{150}$Nd,
and $^{238}$U, the electron shells with probability $\sim 1/4$ turn out to be the lowest excited
state with quantum numbers inherited from the parent atom.
Such decays are accompanied by a subsequent
de-excitation with characteristic emission of photons of the ultraviolet range.
In the atoms $^{48}$Ca, $^{76}$Ge, $^{116}$Cd, $^{128}$Te, $^{130}$Te, and $^{136}$Xe,
the daughter ion's electrons move to the ground state with a probability of $\sim 1/4$ and to an excited state with a probability of $\sim 3/4$.
The average value and variance of the excitation energy were computed for each of the scenarios under consideration.
The dependence on the atomic number indicates the dominant
contribution to the variance of the Feinberg--Migdal effect.
Deviations of the $\beta$-electrons energy from the decay energy $Q^*$ were estimated for the neutrinoless mode of double-$\beta$ decay.

The work was supported by the grant
\#23-22-00307 of Russian Science Foundation.



\end{document}